\pdfoutput=1
\documentclass[lettersize,journal]{IEEEtran}
\usepackage{amsmath,amsfonts}
\usepackage{array}
\usepackage[caption=false,font=normalsize,labelfont=sf,textfont=sf]{subfig}
\usepackage{textcomp}
\usepackage{stfloats}
\usepackage{url}
\usepackage{verbatim}
\usepackage{graphicx}
\DeclareGraphicsExtensions{.pdf,.png,.jpg}
\usepackage{cite}
\usepackage{braket}
\usepackage{algorithm}
\usepackage{algpseudocode}
\usepackage{booktabs}
\usepackage{booktabs, multirow, siunitx}

\begin{document}

\title{Quantum-Key-Distribution Authenticated Aggregation and Settlement for Virtual Power Plants}

\author{Ziqing Zhu,~\IEEEmembership{Member,~IEEE} 
}
\maketitle
\bstctlcite{BSTcontrol}

\begin{abstract}
The proliferation of distributed energy resources (DERs) and demand-side flexibility has made virtual power plants (VPPs) central to modern grid operation. Yet their end-to-end business pipeline, covering bidding, dispatch, metering, settlement, and archival, forms a tightly coupled cyber–physical–economic system where secure and timely communication is critical. Under the combined stress of sophisticated cyberattacks and extreme weather shocks, conventional cryptography offers limited long-term protection. Quantum key distribution (QKD), with information-theoretic guarantees, is viewed as a gold standard for securing critical infrastructures. However, limited key generation rates, routing capacity, and system overhead render key allocation a pressing challenge: scarce quantum keys must be scheduled across heterogeneous processes to minimize residual risk while maintaining latency guarantees. This paper introduces a quantum-authenticated aggregation and settlement framework for VPPs. We first develop a system–threat model that connects QKD key generation and routing with business-layer security strategies, authentication strength, refresh frequency, and delay constraints, providing upper bounds on residual attack success. Building on this, we formulate a key-budgeted risk minimization problem that jointly accounts for economic risk, service-level violations, and key-budget feasibility, and reveal a threshold property linking marginal security value to shadow prices. This structure allows key allocation to be cast as a fractional knapsack problem with approximation guarantees. Algorithmically, we design a hybrid offline–online scheme: offline pre-allocation uses scenario trees and robust optimization to distribute domain-level quotas, while online rolling control applies proximal-dual updates with incremental adjustments, yielding an interpretable price–threshold policy. Case studies on a representative VPP system, incorporating attack pulses, weather shocks, and market contexts, demonstrate that the proposed approach significantly reduces residual risk and SLA violations, enhances key efficiency and robustness, and aligns observed dynamics with the theoretical shadow price mechanism.
\end{abstract}
\begin{IEEEkeywords}
Virtual Power Plant (VPP), Quantum Key Distribution (QKD), Risk-Aware Scheduling, Cyber--Physical Security, QoSec-Constrained Optimization
\end{IEEEkeywords}

\section{Introduction}\label{sec1}
The rapid proliferation of distributed energy resources (DERs) and demand-side flexibility has made the concept of the \emph{virtual power plant} (VPP) a cornerstone of modern power system operation. By aggregating heterogeneous resources and enabling their participation in electricity markets, VPPs provide both economic and reliability benefits \cite{Chen2024RTVPP,Wang2024PriceMaker,ZHANG2023DRC}. Yet the end-to-end business pipeline of a VPP—spanning bidding and clearing, dispatch and acknowledgment, metering upload, settlement and reconciliation, and archival—creates a tightly coupled ``cyber–physical–economic'' system. Secure and timely communication is indispensable: message integrity, confidentiality, and replay protection directly affect settlement outcomes and compliance costs, while end-to-end latency determines the feasibility of dispatch instructions and the value captured in market transactions \cite{Yi2023VPPASRL}.

At the same time, cyberattacks targeting energy infrastructure are becoming more frequent and sophisticated, and extreme weather events can simultaneously disrupt measurement channels and alter market states. This dual stress causes operational risk and system load to fluctuate in sync, amplifying the consequences of both. Against this backdrop, \emph{quantum key distribution} (QKD) has emerged as a promising solution, offering information-theoretic guarantees for key generation and distribution and showing feasibility in utility settings \cite{Aggarwal2024QKD}. For critical infrastructures such as VPPs, QKD is widely regarded as a gold standard for future-proof communication security; however, deployment faces practical barriers: key generation rates are limited by channel conditions and environment, cross-domain routing is constrained by capacity and policy, and end systems are bounded by processing and bandwidth. The central challenge is thus clear: with scarce quantum keys, how should one allocate them across heterogeneous business processes to minimize residual economic risk while preserving service-level agreements (SLAs) on latency?

Addressing this challenge is non-trivial. VPP traffic classes differ sharply in their security and latency requirements as well as in their economic consequences: metering and settlement messages are highly sensitive to tampering and replay, while bidding and dispatch messages demand ultra-low latency. This necessitates fine-grained selection among alternative cryptographic strategies—ranging from OTP+WC with information-theoretic security, to AES+WC hybrids, to AES+MAC with computational security—and careful adjustment of tag lengths and key-refresh frequencies. Meanwhile, key supply, routing, and consumption are dynamically coupled: QKD yields fluctuate with physical conditions; inter-domain flows face capacity and quota limits; and key pools must manage expiration and revocation. Decision-making therefore depends not only on the present state but also on its temporal evolution. Moreover, adversarial intensity and system context (e.g., peak loads or settlement deadlines) are inherently non-stationary, producing amplified losses in critical periods. Together, these factors create a large-scale mixed-integer, nonconvex optimization problem. Achieving rolling, real-time control requires balancing robustness to uncertainty, computational tractability, and interpretability, while also ensuring feasibility recovery under extreme conditions \cite{Chen2024RTVPP,Wang2024PriceMaker,ZHANG2023DRC,Yi2023VPPASRL}.

This paper makes four main contributions. First, we introduce an end-to-end \emph{system–threat model} that links physical-layer QKD key generation and routing with business-layer strategy choices, authentication strength, refresh rates, and the resulting delay constraints, while providing rigorous upper bounds on residual attack success probabilities. This establishes a causal chain from security to economics and latency. Second, we propose the \emph{quantum-authenticated aggregation and settlement} framework, formulated as a key-budgeted risk minimization problem. The model integrates expected economic risk, SLA violations, and key budget feasibility into a unified optimization, and reveals a structural threshold property between marginal security value (MSV) and shadow prices. Third, we design a hybrid offline–online algorithm: offline pre-allocation leverages scenario trees and robust optimization to distribute domain-level quotas, while online rolling control employs proximal–dual updates with incremental parameter adjustments, yielding an interpretable price–threshold policy. Finally, we implement the framework on a representative VPP test system with multi-source data (attack pulses, weather shocks, and business contexts), establishing an evaluation suite that covers overall performance, resource dynamics, and QoSec/latency compliance for critical classes. Results show that the proposed approach substantially reduces residual risk and SLA violations while improving key efficiency and robustness, and that its behavior aligns with the shadow price–strategy dynamics predicted by the theory.

\section{Related Work}
VPP-related research has evolved from early market-participation and bidding models to risk-aware aggregation and multi-time-scale scheduling. Foundational work on VPP bidding and market integration \cite{Mashhour2011TPWRS, Koraki2018TPWRS} was followed by bi-level and multi-operator formulations that coordinate heterogeneous distributed energy resources (DERs) under uncertainty \cite{Wei2018AE, Kong2019AE}. Recent studies develop robust and distributionally robust policies that co-optimize day-ahead and intraday decisions, represent price and renewable uncertainty, and incorporate learning-based scenario generation \cite{Kong2020AE, Li2024AE, Xiong2024AE, Wang2024AE_Pricing, Ma2025AE}. Comprehensive surveys synthesize operational challenges—forecasting, reserve co-optimization, and multi-energy coupling—highlighting the need for scalable algorithms and reliable cyber–physical coordination \cite{Gao2024AE_Review}. Parallel work quantifies the reliability value of DER portfolios, reinforcing the importance of flexible aggregation for resilience \cite{Wang2019TSG}.

Security for power-system communication has been addressed through standards-driven hardening and latency-aware protocol design. Prior studies analyze limitations of IEC~62351 for substation traffic and propose schemes that balance integrity/authentication with strict real-time constraints \cite{Zhang2019MPCE}, while overviews of PMU/WAMS emphasize timing and trust requirements for wide-area protection and control \cite{Phadke2018MPCE}. Related efforts show that both uncertainty-aware VPP scheduling \cite{Ai2016PCMP} and countermeasures for IEC~61850 attack surfaces \cite{Hussain2023PCMP} materially shape feasible operation regions by imposing cyber constraints. QKD has begun to appear in energy and CPS security via systems work that integrates quantum-derived keys with modern key management. A notable example combines QKD and post-quantum cryptography for smart-grid authentication, illustrating deployment-minded architectures and trust anchors beyond purely computational security \cite{Aggarwal2024TNSM}. Still, most VPP and grid-security papers either assume abundant symmetric keys or treat security as fixed overhead, leaving the \emph{economics of keys}—how to allocate scarce QKD keys across time, nodes, and message classes—largely unexplored.

Against this backdrop, our work differs in two ways. First, we introduce \emph{risk-aware key scheduling} that treats secret keys as a networked commodity with state dynamics and shadow prices, jointly optimizing strategy selection, tag length, and refresh under routing and domain quotas. Second, we impose explicit QoSec (probabilistic security) and latency constraints, tying residual attack success probabilities to per-message cryptographic choices and queueing effects. This bridges robust VPP scheduling \cite{Wei2018AE, Kong2020AE, Li2024AE} with standards-aware power-system security \cite{Zhang2019MPCE, Hussain2023PCMP}, while operationalizing QKD-era key scarcity within an optimization and online-control framework \cite{Aggarwal2024TNSM}.

\section{System \& Threat Model}
\allowdisplaybreaks

We consider a VPP aggregating distributed energy resources, that participates in electricity markets via an aggregator. Control and data exchange use a QKD–enabled network. Time is slotted as \(t\in\{0,1,\dots,T\}\), during which bidding/clearing, dispatch, metering, settlement, and archival occur. To capture heterogeneity in security, latency, and economic impact, messages are classified into metering (M1), bidding (M2), dispatch (M3), settlement (M4), and audit (M5), with \(\mathcal{C}=\{\mathrm{M1},\ldots,\mathrm{M5}\}\). For each class \(i\in\mathcal{C}\), let \(\{A_{i,t}\}\) be the arrival process (Poisson with intensity \(\lambda_i\) or general renewal), \(L_i\) the payload size, \(D_i^{\max}\) the latency bound, and \(\mathcal{L}_i>0\) the unit economic loss from successful tampering or replay (e.g., imbalance penalties, compensation, fines, or reputation loss). End-to-end delay in slot \(t\) combines queueing and cryptographic overheads:
\begin{align}
\mathrm{Delay}_{i,t}(s,a,r)
&=\;W_{i,t}
+\tau^{\mathrm{enc}}_{i,t}(s,a,r)
+\tau^{\mathrm{net}}_{i,t}
+\tau^{\mathrm{ver}}_{i,t}(s,a),
\end{align}
where \(s\in\{1,2,3\}\) is the security strategy, \(a\) the authentication-strength parameter (e.g., tag length), and \(r\) the session-key refresh rate. Here \(W_{i,t}\) is queueing delay, \(\tau^{\mathrm{enc}}_{i,t}\) and \(\tau^{\mathrm{ver}}_{i,t}\) are encryption and verification costs, and \(\tau^{\mathrm{net}}_{i,t}\) is transmission time (including header/tag overhead). Service-level agreements require \(\mathrm{Delay}_{i,t}\le D_i^{\max}\).

\subsection{QKD Key Supply and Routing Dynamics}

Secret key supply is provided by a QKD overlay with quantum links \(\mathcal{E}\). For link \(e\in\mathcal{E}\) and slot \(t\), let \(g_{e,t}\) (bits/slot) denote its secret-key yield, which depends on channel fading, QBER, weather, and routing policy. Abstractly, we map observable environment states into yield via a monotone function \(\psi_e\):
\begin{align}
g_{e,t}
&=\;\psi_e\!\big(Q_{e,t},\;\mathrm{SNR}_{e,t},\;\xi_{e,t}\big),
\end{align}
where \(Q_{e,t}\) is the QBER, \(\mathrm{SNR}_{e,t}\) collects physical-layer quality indicators, and \(\xi_{e,t}\) aggregates environmental features such as temperature/humidity and precipitation/wind; \(\psi_e\) is decreasing in \(Q_{e,t}\) and increasing in \(\mathrm{SNR}_{e,t}\) and link availability. Keys can be routed among network nodes through authenticated classical channels and trusted relays to form ``key flows,'' subject to relay processing limits and administrative policies. Let \(\mathcal{V}\) be the node set, and each node \(u\in\mathcal{V}\) maintains a key pool \(K_{u,t}\). The key-pool dynamics in slot \(t\) obey
\begin{align}
K_{u,t+1}
&=\;\min\!\Big\{K_{u}^{\max},\;
K_{u,t}
+ \underbrace{\sum_{e\in\mathrm{In}(u)} g_{e,t}}_{\text{local generation \& inflow}}
+ \underbrace{\sum_{v\in\mathcal{V}} f_{v\to u,t}}_{\text{routed inflow}}
\nonumber\\
&\quad
- \underbrace{\sum_{v\in\mathcal{V}} f_{u\to v,t}}_{\text{routed outflow}}
- \underbrace{\sum_{i\in\mathcal{C}} k_{i,u,t}}_{\text{business consumption}}
- \delta_{u,t}\Big\}.
\end{align}
where \(K_u^{\max}\) is the capacity cap, \(f_{u\to v,t}\) is the routed key flow from \(u\) to \(v\) in slot \(t\), constrained by link/relay capacity \(\sum_{(x,y)\in \mathcal{P}(e)} f_{x\to y,t}\le g_{e,t}\) (with \(\mathcal{P}(e)\) the set of paths traversing link \(e\)), and \(\delta_{u,t}\) captures key expiration and revocation (e.g., purging keys older than a TTL \(\tau^{\mathrm{ttl}}\)). The consumption term \(k_{i,u,t}\) is the net key usage at node \(u\) for class \(i\) in slot \(t\) under the chosen security strategy, detailed below. This state equation explicitly couples security demand with key supply and yields an optimizable ``state--resource'' interface for budgeting and scheduling.

\subsection{Security Options and Per-Message Key Cost}

To trade off security strength against key expenditure, we offer three mutually exclusive strategy options per message: S1: one-time pad (OTP) encryption + Wegman--Carter (WC) universal-hash authentication (information-theoretic security); S2: symmetric block cipher (AES) encryption + WC authentication (computational confidentiality + information-theoretic authentication); S3: AES encryption + computational MAC (e.g., HMAC/KMAC/CMAC). Let \(x^{(s)}_{i,t}\in\{0,1\}\) indicate whether strategy \(s\in\{1,2,3\}\) is chosen for class \(i\) in slot \(t\), with \(\sum_s x^{(s)}_{i,t}=1\). The WC authentication strength is controlled by the tag length \(\ell_{\mathrm{mac}}(a_{i,t})\), where \(a_{i,t}\in\mathbb{R}_+\) is the ``auth-strength knob''; computational MAC tag length is \(\ell_{\mathrm{tag}}\), and the AES session-key refresh frequency is \(r_{i,t}\in\mathbb{Z}_{\ge 1}\). The per-message key consumption is approximated by
\begin{align}
\kappa^{(1)}_i(a_{i,t})
&=\; L_i + \ell_{\mathrm{mac}}(a_{i,t}), \\[6pt]
\kappa^{(2)}_i(a_{i,t},r_{i,t})
&=\; \ell_{\mathrm{iv}} + \ell_{\mathrm{mac}}(a_{i,t})
   + \frac{\ell_{\mathrm{key}}}{r_{i,t}}, \\[6pt]
\kappa^{(3)}_i(r_{i,t})
&=\; \ell_{\mathrm{iv}} + \ell_{\mathrm{tag}}
   + \frac{\ell_{\mathrm{key}}}{r_{i,t}}.
\end{align}
where \(\ell_{\mathrm{iv}}\) is the IV length, \(\ell_{\mathrm{key}}\) is the per-session key length (refreshing once consumes \(\ell_{\mathrm{key}}\) bits of QKD key), and \(\ell_{\mathrm{mac}}(\cdot)\) can be linear or piecewise-linear to match implementation. Hence, the total business key usage at node \(u\) in slot \(t\) is
\begin{align}
k_{i,u,t}
&=\; \mathbb{E}[A_{i,t}]\cdot \Big(
   x^{(1)}_{i,t}\kappa^{(1)}_i(a_{i,t})
\nonumber\\
&\qquad\qquad
 + x^{(2)}_{i,t}\kappa^{(2)}_i(a_{i,t},r_{i,t})
\nonumber\\
&\qquad\qquad
 + x^{(3)}_{i,t}\kappa^{(3)}_i(r_{i,t})
\Big).
\end{align}
where we use \(\mathbb{E}[A_{i,t}]\approx \lambda_i\) under steady-state arrivals; with realized counts, the expectation can be replaced by a sample sum without changing the analysis.

\subsection{Adversary Capability and Residual Success Probability}

We adopt a ``strong man-in-the-middle'' adversary abstraction: the adversary can fully observe and tamper with classical communications except the quantum channel of QKD (i.e., control arbitrary forwarding nodes and link queues), inject/modify/replay messages, and induce controllable delays, yet cannot break information-theoretic limits imposed by OTP and WC authentication; for AES and computational MACs, capability is bounded by standard computational assumptions (PRP/PRF) and key-refresh policy. Let \(p_{i,t}\in[0,1]\) be the exogenous attack-attempt probability (or intensity), driven jointly by historical threat intelligence, industry incidents, and extreme-weather triggers. Given an attack attempt, the \emph{residual} success probabilities under different strategies are upper-bounded by
\begin{align}
\rho^{(1)}_{i,t}(a_{i,t})
&\;\le\; 2^{-\ell_{\mathrm{mac}}(a_{i,t})} + \epsilon_{\mathrm{impl}}, \\[6pt]
\rho^{(2)}_{i,t}(a_{i,t},r_{i,t})
&\;\le\; 2^{-\ell_{\mathrm{mac}}(a_{i,t})}
   + \mathrm{Adv}^{\mathrm{AES}}_{\mathrm{ind\text{-}cca}}
     (q_{i,t},\tau_{i,t};r_{i,t}), \\[6pt]
\rho^{(3)}_{i,t}(r_{i,t})
&\;\le\; \mathrm{Adv}^{\mathrm{MAC}}_{\mathrm{forg}}
     (q_{i,t},\tau_{i,t}) + 2^{-\ell_{\mathrm{tag}}}.
\end{align}
where \(\epsilon_{\mathrm{impl}}\) captures a small constant headroom for implementation issues (e.g., randomness quality and side channels), and \(\mathrm{Adv}^{\mathrm{AES}}_{\mathrm{ind\text{-}cca}}\) and \(\mathrm{Adv}^{\mathrm{MAC}}_{\mathrm{forg}}\) are advantage functions increasing in attack-query budget \(q_{i,t}\) and attack duration \(\tau_{i,t}\), and decreasing in refresh frequency \(r_{i,t}\) (available either from standard reductions or fitted empirical curves). With OTP+WC, residual success is controlled solely by the WC tag length; with AES+WC, authentication remains information-theoretic while confidentiality is reinforced by larger \(r_{i,t}\) and tighter replay windows; with AES+computational MAC, both dimensions rely on computational advantages and are more sensitive to refresh policy and replay-window configuration.

Because the consequences and exploitable surfaces differ across classes, we model the economic loss of a successful attack as
\begin{align}
\mathrm{Loss}_{i,t}
&=\;\mathcal{L}_i \cdot \mathbf{1}\{\mathrm{Attack~succeeds~on~}i\}\cdot \Theta_{i,t},
\end{align}
where \(\Theta_{i,t}\in[0,1]\) is a contextual amplification factor reflecting marginal harm variations under different system states (e.g., peak load, binding market-clearing constraints, end-of-day settlement windows). The slot-\(t\) \emph{expected residual economic risk} is therefore
\begin{align}
\mathbb{E}[\mathrm{Risk}_t]
&=\;\sum_{i\in\mathcal{C}} p_{i,t}\;
\Big(
   x^{(1)}_{i,t}\rho^{(1)}_{i,t}(a_{i,t})
\nonumber\\
&\qquad\qquad
 + x^{(2)}_{i,t}\rho^{(2)}_{i,t}(a_{i,t},r_{i,t})
\nonumber\\
&\qquad\qquad
 + x^{(3)}_{i,t}\rho^{(3)}_{i,t}(r_{i,t})
\Big)
\nonumber\\
&\qquad\qquad\times\;
\mathcal{L}_i \;\mathbb{E}[\Theta_{i,t}].
\end{align}
which provides a (piecewise) differentiable mapping from ``strategy selection/auth-strength/refresh rate/key consumption'' to ``residual risk,'' forming the central bridge for key-budget optimization.

\subsection{Latency Constraints and Queueing Approximation}

End-to-end latency constraints couple security-induced expansion and computation costs with available bandwidth and queue occupancy. Let the effective link bandwidth be \(B_{t}\) (bits/slot), so the serialization time per message of class \(i\) is \((L_i+\Delta L_i(s,a))/B_{t}\), where \(\Delta L_i(s,a)\) is overhead induced by headers, tags, and nonces under strategy \((s,a)\). Using the Kingman approximation for a GI/G/1 queue, we have
\begin{align}
W_{i,t}
&\;\approx\;\frac{\rho_t}{1-\rho_t}\cdot \frac{c_a^2 + c_s^2}{2}\cdot \frac{1}{\mu_{i,t}(s,a,r)},
\\
\rho_t
&=\sum_{i}\frac{\lambda_i}{\mu_{i,t}(s,a,r)},
\end{align}
where \(c_a^2\) and \(c_s^2\) are the squared coefficients of variation of inter-arrival and service times, and \(\mu_{i,t}^{-1}(s,a,r)\) absorbs mean crypto (enc/auth and verification) time as well as transmission and retransmission overhead. This approximation enables rapid design-time screening of \((s,a,r)\) effects on delay and is enforced via hard/soft constraints \(\mathrm{Delay}_{i,t}\le D_i^{\max}\) (with timeout penalties).

\subsection{Domain-Level Key-Flow Constraints and Summary}

To reflect topology and inter-domain key-transit realities, we impose domain-level caps \(B_{d,t}^{\mathrm{key}}\) for any management domain \(d\) and slot \(t\):
\begin{align}
\sum_{(u,v)\in \mathcal{E}_d} f_{u\to v,t}
&\;\le\; B_{d,t}^{\mathrm{key}},\\
\sum_{i\in\mathcal{C}} k_{i,u,t}^{(d)}
&\;\le\;K^{\mathrm{alloc}}_{d,t},
\end{align}
where \(\mathcal{E}_d\) collects intra-domain relay links and \(K^{\mathrm{alloc}}_{d,t}\) is the domain-level allocable key quota. These constraints render the budgeting problem spatially a multi-commodity flow and align with the geographic distribution and priority of business traffic. In summary, this section provides a unified system--threat model from physical-layer key generation and routing, to business-layer strategy selection and delay constraints, and further to adversarial advantage and residual risk. The key state is the node key pools \(\{K_{u,t}\}\); the key controls are \((x^{(s)}_{i,t},a_{i,t},r_{i,t})\); and the key costs are \(\mathbb{E}[\mathrm{Risk}_t]\) and latency-violation penalties. The model captures hybrid information-theoretic and computational security while preserving fine-grained engineering facets (refresh, routing, bandwidth, expiration), offering a rigorous and computable foundation for subsequent key-budgeted risk minimization and rolling online scheduling.

\section{Key-Budgeted Risk Minimization}
\allowdisplaybreaks

Building upon the system--threat characterization in the previous section, we now formalize the \emph{key-budgeted risk minimization} problem. Over discrete slots \(t=0,1,\dots,T\), we jointly decide, for each class \(i\in\mathcal{C}\), the security strategy \(x_{i,t}^{(s)}\in\{0,1\}\) with \(s\in\{1,2,3\}\) and \(\sum_s x_{i,t}^{(s)}=1\), the authentication-strength control \(a_{i,t}\ge 0\) (determining the WC-MAC tag length \(\ell_{\mathrm{mac}}(a_{i,t})\)), and the session-key refresh frequency \(r_{i,t}\in\mathbb{Z}_{\ge 1}\). These are coupled with key-routing flows \(f_{u\to v,t}\) and node key-pool dynamics \(K_{u,t}\) to minimize a weighted cumulative cost that accounts for residual economic risk, latency violations, and infeasible key budgets. Let \(\rho_{i,t}^{(s)}(a_{i,t},r_{i,t})\) denote the residual-risk mapping from the previous section, \(k_{i,u,t}(x,a,r)\) the key consumption, and \(\mathrm{Delay}_{i,t}(s,a,r)\) the end-to-end latency. We use the positive-part operator \([z]_+ := \max\{z,0\}\) and the indicator \(\mathbf{1}\{\cdot\}\).

\subsection{Objective}
\allowdisplaybreaks

We seek a policy that trades off (i) expected residual economic risk from successful attacks, (ii) soft penalties for end-to-end latency violations, (iii) soft penalties for temporary key-budget infeasibility (to discourage over-consumption of keys), and (iv) a smoothing term that penalizes rapid switching of strategies or aggressive retuning of authentication strength and refresh rates. Formally, we minimize
\begin{align}
J
&= \sum_{t=0}^{T} \bigg\{
\mathbb{E}[\mathrm{Risk}_t]
+ \sum_{i\in\mathcal{C}} \phi_i \,
  \big[\mathrm{Delay}_{i,t}(s,a,r) - D_i^{\max}\big]_+
\nonumber\\
&\qquad
+ \eta \bigg[
  \sum_{u\in\mathcal{V}} \sum_{i\in\mathcal{C}}
    k_{i,u,t}(x,a,r)
\nonumber\\
&\qquad\qquad
  - \sum_{u\in\mathcal{V}}
    \Big( K_{u,t}
    + \sum_{e\in\mathrm{In}(u)} g_{e,t} \Big)
\bigg]_+
\nonumber\\
&\qquad
+ \varpi \,\Xi_t(x,a,r)
\bigg\}.
\end{align}
The first term aggregates residual risk in slot \(t\), weighted by the business loss parameters; the second adds a per-class SLA penalty for any excess latency; the third applies a hinge penalty whenever instantaneous key demand exceeds locally available key stock and inflow; and the last promotes temporal smoothness to avoid churning implementations and control oscillations.

The expected residual risk in slot \(t\) aggregates, across classes, the attack attempt probability \(p_{i,t}\), the class-specific residual success probability determined by the chosen security option, and the class loss \(\mathcal{L}_i\) scaled by a context factor:
\begin{align}
\mathbb{E}[\mathrm{Risk}_t]
&=\sum_{i\in\mathcal{C}} p_{i,t}\,
\Big(
   x_{i,t}^{(1)} \rho_{i,t}^{(1)}(a_{i,t})
\nonumber\\
&\qquad\qquad
 + x_{i,t}^{(2)} \rho_{i,t}^{(2)}(a_{i,t},r_{i,t})
 + x_{i,t}^{(3)} \rho_{i,t}^{(3)}(r_{i,t})
\Big)\,
\mathcal{L}_i\,\mathbb{E}[\Theta_{i,t}].
\end{align}
Here \(\rho^{(s)}\) is the residual success bound under strategy \(s\) (defined precisely below), and \(\Theta_{i,t}\in[0,1]\) captures how current operating context amplifies loss (e.g., peak settlement windows). The SLA penalty weights \(\phi_i>0\) encode the relative urgency of latency per class. The coefficient \(\eta>0\) sets how strongly we discourage using more keys than available in the current slot (a soft budget), while \(\varpi\ge 0\) weights the smoothing term
\begin{align}
\Xi_t(x,a,r)
&=\sum_{i\in\mathcal{C}} \bigg(
  \zeta_a \,|a_{i,t}-a_{i,t-1}|
\nonumber\\
&\qquad\qquad
+ \zeta_r \,|r_{i,t}-r_{i,t-1}|
\nonumber\\
&\qquad\qquad
+ \zeta_x \sum_{s} |x_{i,t}^{(s)}-x_{i,t-1}^{(s)}|
\bigg),
\end{align}
where \(\zeta_a,\zeta_r,\zeta_x\ge 0\) discourage abrupt changes of authentication strength \(a\), refresh rate \(r\), and strategy choices \(x\), respectively. In receding-horizon implementations, we restrict the sum to a short window \(t,\dots,t+H-1\) and append a terminal potential \(V_{t+H}(K_{\cdot,t+H})\) to capture the future value of remaining keys, thereby balancing near-term feasibility with long-term prudence.

\subsection{Constraints}

\paragraph{Key-pool and routing constraints (state evolution and capacities).}
Keys are produced by QKD links, routed through trusted relays, stored in node key pools, and consumed by business traffic according to selected strategies. The key-pool state for node \(u\) evolves as
\begin{align}
K_{u,t+1}
&=\; \min\!\Big\{
K_u^{\max},\;
K_{u,t}
\nonumber\\
&\qquad
+ \sum_{e\in\mathrm{In}(u)} g_{e,t}
+ \sum_{v\in\mathcal{V}} f_{v\to u,t}
\nonumber\\
&\qquad
- \sum_{v\in\mathcal{V}} f_{u\to v,t}
- \sum_{i\in\mathcal{C}} k_{i,u,t}(x,a,r)
- \delta_{u,t}
\Big\},
\end{align}
where \(g_{e,t}\) is the QKD yield on inbound links to \(u\), \(f_{v\to u,t}\) and \(f_{u\to v,t}\) are routed inflow/outflow, \(k_{i,u,t}\) is business consumption induced by \((x,a,r)\), and \(\delta_{u,t}\) models expirations/revocations. Feasibility requires nonnegativity and capacity/quota compliance:
\begin{align}
K_{u,t} &\ge 0,\qquad f_{u\to v,t}\ge 0,
\\
\sum_{(x,y)\in\mathcal{E}_d} f_{x\to y,t}
&\le B_{d,t}^{\mathrm{key}},\qquad
\sum_{i} k_{i,u,t}^{(d)} \le K_{d,t}^{\mathrm{alloc}},
\\
\sum_{(x,y)\in\mathcal{P}(e)} f_{x\to y,t}
&\le g_{e,t}.
\end{align}
The first line enforces physical nonnegativity; the second aggregates per-domain transit and allocable quotas; the third caps any path set traversing a QKD link \(e\) by its yield.

\paragraph{Service and compliance constraints (latency and minimum security).}
End-to-end latency must respect SLA bounds, possibly softened in the objective:
\begin{align}
\mathrm{Delay}_{i,t}(s,a,r) \;\le\; D_i^{\max}.
\end{align}
For critical classes (e.g., M1 metering, M4 settlement), we forbid weak options and enforce minimum tag strength:
\begin{align}
x_{i,t}^{(3)}=0,\qquad \ell_{\mathrm{mac}}(a_{i,t}) \;\ge\; \ell_{\min}.
\end{align}

\paragraph{Feasible strategy domain.}
Choices are restricted to the discrete/boxed domain
\begin{align}
x_{i,t}^{(s)} &\in \{0,1\}, \qquad 
   \sum_s x_{i,t}^{(s)} = 1, \\[6pt]
a_{i,t} &\in [0,a_{\max}], \\[6pt]
r_{i,t} &\in \{1,2,\dots,r_{\max}\}.
\end{align}

\paragraph{Structural assumptions for computation (monotonicity/convexification aids).}
To enable convex relaxations and efficient online control, we assume the residual success bounds behave monotonically with respect to design knobs:
\begin{align}
\rho_{i,t}^{(1)}(a)
&= 2^{-\ell_{\mathrm{mac}}(a)} + \epsilon_{\mathrm{impl}}, \\[6pt]
\rho_{i,t}^{(2)}(a,r)
&= 2^{-\ell_{\mathrm{mac}}(a)}
   + \mathrm{Adv}^{\mathrm{AES}}_{\mathrm{ind\text{-}cca}}(q_{i,t},\tau_{i,t};r), \\[6pt]
\rho_{i,t}^{(3)}(r)
&= \mathrm{Adv}^{\mathrm{MAC}}_{\mathrm{forg}}(q_{i,t},\tau_{i,t})
   + 2^{-\ell_{\mathrm{tag}}}.
\end{align}

Here, $\rho^{(1)}$ decreases in $a$ (longer WC tags reduce forgery probability, up to an implementation headroom $\epsilon_{\mathrm{impl}}$).  
$\rho^{(2)}$ decreases in both $a$ and $r$ (stronger authentication and more frequent refresh both help).  
$\rho^{(3)}$ decreases in $r$ (computational MAC forgery bound plus a fixed tag term).
Per-message key costs grow with security strength: \(\kappa_i^{(1)}\) increases with \(a\) (WC tag bits); \(\kappa_i^{(2)}\) increases with \(a\) and with \(1/r\) (more frequent session-key use); \(\kappa_i^{(3)}\) increases with \(1/r\) (computational MAC tag fixed, but refresh still consumes QKD key). Consequently, the expected consumption for class \(i\) at node \(u\) in slot \(t\) is
\begin{align}
k_{i,u,t}
&=\; \mathbb{E}[A_{i,t}] \cdot \Big(
   x^{(1)}_{i,t}\kappa^{(1)}_i(a_{i,t})
\nonumber\\
&\qquad\qquad
 + x^{(2)}_{i,t}\kappa^{(2)}_i(a_{i,t},r_{i,t})
\nonumber\\
&\qquad\qquad
 + x^{(3)}_{i,t}\kappa^{(3)}_i(r_{i,t})
\Big),
\end{align}
where \(\mathbb{E}[A_{i,t}]\approx \lambda_i\) under steady-state arrivals (or replaced by realized counts in implementation). This closes the loop between strategy choices \((x,a,r)\), residual success probabilities \(\rho\), latency \(\mathrm{Delay}\), and key consumption \(k\), making the resource–risk–latency trade-offs explicit and amenable to convexification and online dual-based control.

\subsection{Computational Relaxations}

Because of binary \(x\) and discrete \(r\), the original problem is a large-scale mixed-integer nonconvex program. For day-ahead/day-of pre-allocation, we adopt a two-step convexification. First, introduce a fractional selection \(y_{i,t}^{(s)}\in[0,1]\) for the proportion of class-\(i\) messages using strategy \(s\) in slot \(t\), replacing \(\sum_s y_{i,t}^{(s)}=1\) and rewriting
\begin{align}
k_{i,u,t}(x,a,r)
&\;\leadsto\;
\lambda_i \Big(
  y_{i,t}^{(1)} \kappa_i^{(1)}(a_{i,t})
\nonumber\\
&\qquad\qquad
+ y_{i,t}^{(2)} \kappa_i^{(2)}(a_{i,t},r_{i,t})
\nonumber\\
&\qquad\qquad
+ y_{i,t}^{(3)} \kappa_i^{(3)}(r_{i,t})
\Big).
\end{align}
Second, approximate the nonlinearities in \(\rho\), \(\kappa\), and \(\mathrm{Delay}\) by piecewise-convex upper bounds (e.g., using breakpoints of \(a\) to piecewise-linearize \(\ell_{\mathrm{mac}}(a)\), and discrete points of \(1/r\) with perspective constraints), yielding an MICP/MISOCP with linear or second-order cone constraints. For rolling online decisions, within a short horizon \(H\), one may fix a candidate set for \(y\) (e.g., the previous solution and local variants), optimize only the continuous \((a,r)\), and then quantize \(y\) back to \(\{0,1\}\) heuristically for strategy assignment to meet real-time requirements.

\subsection{Lagrangian Relaxation and Marginal Security Value}

To reveal where ``each bit of key is most valuable,'' we apply Lagrangian relaxation, absorbing cross-node and cross-domain key constraints into the objective with dual multipliers (shadow prices) \(\pi_{u,t}\ge 0\), \(\pi_{d,t}^{\mathrm{key}}\ge 0\), and \(\pi_t^{\mathrm{pool}}\ge 0\), and form
\begin{align}
\mathcal{L}
&=\; \sum_{t} \Big\{
\mathbb{E}[\mathrm{Risk}_t]
+ \sum_{i} \phi_i \big[\mathrm{Delay}_{i,t} - D_i^{\max}\big]_+
+ \varpi \,\Xi_t \Big\}
\nonumber\\
&\quad+\; \sum_{t,u} \pi_{u,t}
\Big( \sum_{i} k_{i,u,t} - K_{u,t}
- \sum_{e\in\mathrm{In}(u)} g_{e,t} \Big)
\nonumber\\
&\quad+\; \sum_{t,d} \pi_{d,t}^{\mathrm{key}}
\Big( \sum_{(x,y)\in\mathcal{E}_d} f_{x\to y,t}
- B_{d,t}^{\mathrm{key}} \Big)
\nonumber\\
&\quad+\; \pi_t^{\mathrm{pool}} \Big(
\sum_{u,i} k_{i,u,t}
- \sum_{u} \big(K_{u,t} + \sum_{e\in\mathrm{In}(u)} g_{e,t}\big)
\Big).
\end{align}
Given dual prices, the class-wise choice of \((s,a,r)\) reduces to a pointwise trade-off between ``marginal risk reduction per key bit'' and shadow price. Let \(\Delta \kappa_i^{(s)}\) denote the extra key consumption when moving from a weaker to a stronger strategy/parameter, and \(\Delta \rho_i^{(s)}(a,r)\) the corresponding drop in residual success probability. We define the \emph{marginal security value} (MSV) as
\begin{align}
\mathrm{MSV}_{i,t}^{(s)}
=\; \frac{
p_{i,t}\, \Delta \rho_i^{(s)}(a,r)\, \mathcal{L}_i \,\mathbb{E}[\Theta_{i,t}]
}{
\Delta \kappa_i^{(s)}
}.
\end{align}
KKT conditions imply that, when latency terms are inactive or negligible, if \(\mathrm{MSV}_{i,t}^{(s)} > \bar{\pi}_t\) (an appropriately aggregated shadow price, e.g., a weighted average across nodes/domains), the optimizer prefers a stronger strategy or higher \(a\), \(r\); if \(\mathrm{MSV}_{i,t}^{(s)} < \bar{\pi}_t\), it prefers downgrading or reducing \(a\), \(r\). More concretely, fixing \(t\) and relaxing \(y_{i,t}^{(s)}\in[0,1]\) with piecewise-linear convex approximations of \(\kappa\) and \(\rho\), the per-slot subproblem over \(\{y_{i,t}^{(s)}\}\) is equivalent to a \emph{fractional knapsack}: allocate stronger protection in descending order of \(\mathrm{MSV}\) until the key budget is met or the balance point \(\mathrm{MSV}=\bar{\pi}_t\) is reached; the remainder adopts next-best strategies. This structure justifies a greedy sorting algorithm with \(O(|\mathcal{C}|\log|\mathcal{C}|)\) complexity per slot.

\subsection{Dynamic Coupling and Online Dual Updates}

Dynamic coupling arises through the key-pool state \(K_{\cdot,t}\). Let \(V_t(K_{\cdot,t})\) be the optimal cost-to-go, satisfying a Bellman-type recursion
\begin{align}
V_t(K)
&=\; \min_{x,a,r,f} \Big\{
\mathbb{E}[\mathrm{Risk}_t]
\nonumber\\
&\qquad
+ \sum_{i} \phi_i \big[\mathrm{Delay}_{i,t}-D_i^{\max}\big]_+
\nonumber\\
&\qquad
+ \varpi\, \Xi_t
\nonumber\\
&\qquad
+ \mathbb{E}\big[ V_{t+1}(K') \big]
\Big\}.
\end{align}
with \(K'\) given by the state equation. Solving this DP exactly is intractable, but subgradient updates of dual prices \(\pi\) approximate the marginal value of key resources:
\begin{align}
\pi_{u,t}^{(n+1)}
=\; \Big[
\pi_{u,t}^{(n)}
+ \gamma_n \Big(
\sum_{i} k_{i,u,t} - K_{u,t}
- \sum_{e\in\mathrm{In}(u)} g_{e,t}
\Big)
\Big]_+,
\end{align}
with stepsizes \(\gamma_n\) satisfying Robbins--Monro conditions. Under statistically stationary or slowly varying \(g_{e,t}\), \(p_{i,t}\), this online update converges to a near-optimal solution; during extreme-weather events that sharply reduce \(g_{e,t}\), \(\pi\) increases (``key shadow price'' rises) to prioritize high-value classes such as M4/M1.

\subsection{Robust/Stochastic Extensions and Feasibility Recovery}

To balance feasibility and robustness, we allow two common extensions. (i) \emph{Uncertainty sets}: introduce a set \(\mathcal{U}_t\) for \((g_{e,t},p_{i,t},\lambda_i)\), e.g., polyhedral or \(\phi\)-divergence balls, and enforce key, delay, and risk constraints for all \((g,p,\lambda)\in\mathcal{U}_t\), or include a worst-case expectation \(\sup_{(g,p,\lambda)\in\mathcal{U}_t}\mathbb{E}[\mathrm{Risk}_t]\) in the objective. (ii) \emph{Chance constraints}: require \(\Pr(K_{u,t}\ge 0)\ge 1-\epsilon_{\mathrm{key}}\) and \(\Pr(\mathrm{Delay}_{i,t}\le D_i^{\max})\ge 1-\epsilon_i\), then convert via Cantelli or Chebyshev bounds into SOCP constraints. In practice, a scenario tree \(\{\omega\in\Omega\}\) with weights \(\pi_\omega\) can be used, writing objectives and constraints as \(\sum_{\omega} \pi_\omega (\cdot)_\omega\) and updating scenario weights in a receding horizon.

The framework naturally accommodates ``hard compliance + soft budget.'' For example, for M4 (settlement) we enforce \(x_{i,t}^{(1)}+x_{i,t}^{(2)}=1\) and \(\ell_{\mathrm{mac}}(a_{i,t})\ge \ell_{\min}\); feasibility can be restored by sacrificing low-priority classes (reducing \(a\) or switching them to S3). For M1 (metering), an explicit QoSec constraint \(\rho_{i,t}^{(s)}(\cdot)\le \epsilon_{\mathrm{meter}}\) can be imposed. If feasibility is still violated, we trigger a \emph{feasibility recovery} subproblem:
\begin{align}
\min_{\{\zeta_i\}}
&\;\sum_{i} \omega_i \,\zeta_i
\nonumber\\
\text{s.t.}\quad
&\text{key and compliance constraints
under relaxations } \zeta_i.
\end{align}
where \(\zeta_i\) quantify relaxation magnitudes (e.g., reducing reporting frequency, aggregating messages, deferring logs) and \(\omega_i\) encode business priorities, ensuring the system degrades to a safe feasible operating point at minimal cost.

\section{Algorithm Design}
\allowdisplaybreaks

This section presents an integrated solution strategy for the QAAS framework combining a \emph{slow timescale} (day-ahead/intra-day planning) to obtain high-quality \emph{key--policy pre-allocation} and \emph{routing/quotas} via scenario-based convexified models, with a \emph{fast timescale} (minute-/second-level rolling control) that performs shadow-price-driven threshold--greedy decisions and small-step proximal updates for real-time feasibility and near-optimality under uncertain key yields and attack intensities.

\subsection{Offline Stage: Scenario MICP with Column Generation and Decomposition}

On an offline horizon \(\mathcal{T}_\mathrm{off}\), we construct a scenario tree \(\Omega\) (from weather--QBER forecasts and threat intelligence) to model \(g_{e,t}\), \(p_{i,t}\), and \(\lambda_i\), and minimize a scenario-weighted expected objective via sample-average approximation. For computability, each class \(i\) uses a finite grid \(A_i=\{a^{(1)},\dots,a^{(M)}\}\) and \(R_i=\{r^{(1)},\dots,r^{(N)}\}\), and we encode each \emph{strategy--parameter} pair as a finite column set \(\mathcal{S}_i=\{(s,a^{(m)},r^{(n)})\}\). Let \(y_{i,t,\omega}^{(s,m,n)}\in[0,1]\) be the fraction of class-\(i\) messages in scenario \(\omega\), slot \(t\), using column \((s,m,n)\), with \(\sum_{s,m,n}y_{i,t,\omega}^{(s,m,n)}=1\). The induced key consumption and residual risk are
\begin{align}
k_{i,u,t,\omega}
&= \lambda_{i,\omega}\!
\sum_{s,m,n} y_{i,t,\omega}^{(s,m,n)}\,
\kappa_i^{(s)}\!\big(a^{(m)},r^{(n)}\big),
\\[6pt]
\mathbb{E}[\mathrm{Risk}_{t,\omega}]
&= \sum_{i\in\mathcal{C}} p_{i,t,\omega}\!
\sum_{s,m,n} y_{i,t,\omega}^{(s,m,n)}\,
\rho_{i,t,\omega}^{(s)}\!\big(a^{(m)},r^{(n)}\big)
\nonumber\\
&\qquad\qquad \times\;
\mathcal{L}_i \,\mathbb{E}[\Theta_{i,t,\omega}].
\end{align}
and Kingman-based service-rate bounds with header inflation yield an SOCP approximation of \(\mathrm{Delay}_{i,t,\omega}\), so latency enters as convex constraints. To avoid enumerating all columns, we employ a \emph{master + pricing (column generation)} scheme. The master problem, with active columns \(\mathcal{S}_i^\mathrm{act}\subseteq \mathcal{S}_i\), solves a MISOCP/MICP and produces duals, notably node/domain key shadow prices \(\pi_{u,t,\omega}\) and latency duals \(\mu_{i,t,\omega}\). The pricing subproblem searches, for each \((i,t,\omega)\), a column \((s^\star,a^{(m)},r^{(n)})\) with positive \emph{reduced profit}
\begin{align}
\Delta \Phi_{i,t,\omega}^{(s,m,n)}
&=\underbrace{p_{i,t,\omega}\!
\left(\rho_{i,t,\omega}^{(\mathrm{base})}
-\rho_{i,t,\omega}^{(s)}(a^{(m)},r^{(n)})\right)\!
\mathcal{L}_i\,\mathbb{E}[\Theta_{i,t,\omega}]}_{\text{benefit from risk reduction}}
\nonumber\\
&\quad-\;
\underbrace{\sum_{u}\bar\pi_{u,t,\omega}\,
\kappa_i^{(s)}(a^{(m)},r^{(n)})}_{\text{cost at key shadow prices}}
\nonumber\\
&\quad-\;
\underbrace{\bar\mu_{i,t,\omega}\,
\Delta \mathrm{Delay}_{i,t,\omega}^{(s,m,n)}}_{\text{latency dual cost}}.
\end{align}
where \(\bar\pi,\bar\mu\) are aggregated from master duals via business--routing mappings. If \(\max_{s,m,n}\Delta \Phi_{i,t,\omega}^{(s,m,n)}\le 0\), the column set is complete. The pricing step is computed by \emph{grid scan + local continuous refinement}: evaluate on \(A_i\times R_i\), then refine \(a\) along one dimension so that the WC tag length meets a first-order balance. For S1 with differentiable \(\ell_{\mathrm{mac}}(a)\), since \(\rho^{(1)}(a)=2^{-\ell_{\mathrm{mac}}(a)}+\epsilon_{\mathrm{impl}}\),
\begin{align}
\frac{\partial}{\partial a}\rho^{(1)}(a)
= -(\ln 2)\,2^{-\ell_{\mathrm{mac}}(a)}\,\ell_{\mathrm{mac}}'(a),
\end{align}
and the reduced-cost stationarity around
\begin{align}
& p_{i,t,\omega}\,\mathcal{L}_i\,\mathbb{E}[\Theta_{i,t,\omega}]\,
\frac{\partial}{\partial a}\rho^{(1)}(a)
\nonumber\\
&\;\;\approx\;
\sum_u \bar\pi_{u,t,\omega}\,
\frac{\partial}{\partial a}\kappa_i^{(1)}(a)
\nonumber\\
&\;\;\quad
+ \bar\mu_{i,t,\omega}\,
\frac{\partial}{\partial a}
   \Delta \mathrm{Delay}_{i,t,\omega}^{(1)}(a).
\end{align}
is reached via Newton/secant steps. Key routing is decoupled from business assignment: the master produces node/domain net demands \(d_{u,t,\omega}\), and a routing subproblem over the QKD topology solves
\begin{align}
\min_{\{f_{x\to y,t,\omega}\ge 0\}} \quad & 0
\nonumber\\
\text{s.t.}\quad
& \sum_{(x,y)\in\mathcal{P}(e)} f_{x\to y,t,\omega}
   \;\le\; g_{e,t,\omega},
\nonumber\\
& \sum_{v} f_{v\to u,t,\omega}
  - \sum_{v} f_{u\to v,t,\omega}
   \;\ge\; d_{u,t,\omega}.
\end{align}
whose feasibility violations generate Benders cuts through \(\pi_{u,t,\omega}\) back to the master. The overall loop nests column generation with Benders cuts, and typically converges in dozens of rounds to a publishable day-ahead plan.

\subsection{Online Stage: Receding Horizon with Threshold--Proximal Refinement}

In real time, at each slot \(t\) we solve a small rolling-horizon (\(H\)) convexified subproblem using the observed \(K_{u,t}\) and short-term forecasts \(\{\hat g_{e,\tau},\hat p_{i,\tau},\hat \lambda_{i,\tau}\}_{\tau=t}^{t+H-1}\), producing feasible near-optimal controls under limited iterations. We fix a candidate column set (offline-optimal columns plus local perturbations), optimize only continuous parameters \((a_{i,\tau},r_{i,\tau})\) and routing flows \(f_{\cdot\to\cdot,\tau}\), and replace full convergence with \emph{one or few} dual steps. Given current duals \(\pi_\tau\), define the proximal augmented Lagrangian
\begin{align}
\mathcal{L}_\mathrm{prox}
&=\sum_{\tau=t}^{t+H-1}\Big\{
\mathbb{E}[\mathrm{Risk}_\tau]
+ \sum_{i}\phi_i\,[\mathrm{Delay}_{i,\tau}-D_i^{\max}]_+
\nonumber\\
&\qquad
+ \varpi\,\Xi_\tau
+ \sum_{u}\pi_{u,\tau}\Big(
     \sum_{i}k_{i,u,\tau}
     - K_{u,\tau}
     - \!\!\sum_{e\in\mathrm{In}(u)}\!\!\hat g_{e,\tau}
   \Big)
\Big\}
\nonumber\\
&\quad
+ \frac{\beta_a}{2}\sum_{i,\tau}\big(a_{i,\tau}-a_{i,\tau-1}\big)^2
\nonumber\\
&\quad
+ \frac{\beta_r}{2}\sum_{i,\tau}\big(r_{i,\tau}-r_{i,\tau-1}\big)^2.
\end{align}
where proximal terms stabilize iteration and suppress jitter. Continuous parameters are updated by projected proximal subgradients; for \(a\) under S1/S2,
\begin{align}
a_{i,\tau}^{(k+1)}
&=\Pi_{[0,a_{\max}]}\!\Big(
a_{i,\tau}^{(k)}
-\eta_k\big[
\nonumber\\
&\qquad
\underbrace{p_{i,\tau}\,\mathcal{L}_i\,\mathbb{E}[\Theta_{i,\tau}]\,
\partial_{a}\rho^{(s)}_{i,\tau}(a)}_{\text{risk gradient}}
\nonumber\\
&\qquad
+\underbrace{\sum_{u}\pi_{u,\tau}\,
\partial_{a}\kappa_i^{(s)}(a)}_{\text{key-cost gradient}}
\nonumber\\
&\qquad
+\underbrace{\phi_i\,
\partial_{a}[\mathrm{Delay}_{i,\tau}-D_i^{\max}]_+}_{\text{latency-penalty gradient}}
\nonumber\\
&\qquad
+\beta_a\,(a_{i,\tau}-a_{i,\tau-1})
\big]\Big).
\end{align}
where \(\partial_{a}\rho^{(1)}=-\ln 2\cdot 2^{-\ell_{\mathrm{mac}}(a)}\,\ell'_{\mathrm{mac}}(a)\); \(\partial_{a}\rho^{(2)}\) is analogous with an additional AES term (negligible or empirically fitted); S3 has no WC so \(\partial_{a}\rho^{(3)}=0\). Since \(r\) is discrete, we use \emph{coordinate search/few-candidate comparison}: for each \(i,\tau\),
\begin{align}
r_{i,\tau}^{\star}
&=\arg\min_{r\in R_i}\Big\{
p_{i,\tau}\,\mathcal{L}_i\,\mathbb{E}[\Theta_{i,\tau}]\,
\rho_{i,\tau}^{(s)}(a,r)
\nonumber\\
&\quad+\sum_{u}\pi_{u,\tau}\,\kappa_i^{(s)}(a,r)
+\phi_i\,[\mathrm{Delay}_{i,\tau}(a,r)-D_i^{\max}]_+
\nonumber\\
&\quad+\frac{\beta_r}{2}\,(r-r_{i,\tau-1})^2
\Big\}.
\end{align}
which costs only a constant factor proportional to \(|R_i|\). Strategy selection follows the real-time \emph{\(\mathrm{MSV}\)-threshold} rule: with current \(\bar\pi_\tau\),
\begin{align}
\mathrm{MSV}_{i,\tau}^{(s)}
=\frac{p_{i,\tau}\,\Delta \rho_{i,\tau}^{(s)}\,\mathcal{L}_i\,\mathbb{E}[\Theta_{i,\tau}]}
{\Delta \kappa_{i}^{(s)}},
\end{align}
and protection is allocated in descending order until the predicted budget \(\hat K_\tau\) (or a proximal dual balance) is met. Duals are updated with a single projected subgradient step,
\begin{align}
\pi_{u,\tau}^{+}
=\Big[\pi_{u,\tau}
+\gamma\Big(\sum_{i}k_{i,u,\tau}-K_{u,\tau}-\!\!\sum_{e\in\mathrm{In}(u)}\!\!\hat g_{e,\tau}\Big)\Big]_+,
\end{align}
then carried as a warm start to \(\tau\!+\!1\) together with \((a,r)\). Under tight compute budgets, the loop degrades to a single pass of “sorting + one proximal step on continuous parameters + one dual update,” which remains feasible and robust due to the threshold structure.

The online loop embeds \emph{adaptive risk calibration} and \emph{exploration--exploitation}. For each class \(i\), maintain a \(\mathrm{Beta}(\alpha_i,\beta_i)\) prior and update it with Bernoulli outcomes from detected compromises/near-misses:
\begin{align}
\hat p_{i,t}
&=\frac{\alpha_i}{\alpha_i+\beta_i},\qquad
\alpha_i\leftarrow \alpha_i
+ \text{(\# detected successful attacks)},
\nonumber\\
\beta_i&\leftarrow \beta_i
+ \text{(\# near-miss/normal events)}.
\end{align}
When uncertainty is large, reserve a fraction \(\beta\in(0,1)\) of an \emph{exploration budget} to momentarily raise protection, effectively replacing \(\hat p_{i,t}\) by a lower-confidence bound in \(\mathrm{MSV}\).

\subsection{Complexity, Implementation, and Robustness Details}

The offline master--pricing--routing loop is dominated by the MISOCP master and pricing scans. With \(|\mathcal{C}|=C\), number of active columns \(Q\), edges \(|\mathcal{E}|=E\), scenarios \(|\Omega|=S\), a typical master iteration empirically scales like \(\tilde O(S\,Q^{1.5})\), pricing like \(O(S\,C\,|A|\,|R|)\) plus constant-step refinements, and routing like \(O(S\,E)\) for linear feasibility/shortest augmenting flows. Online per-slot cost is \(O(C\log C)\) for sorting, \(O(C(|A|+|R|))\) for proximal/coordinate updates, and \(O(|\mathcal{V}|+E)\) for one dual step, well within ms--s times. In practice, function values/derivatives of \(\kappa,\rho,\mathrm{Delay}\) on \((a,r)\) grids are \emph{precomputed and cached}, so online uses table lookups/interpolation. The switching penalty \(\Xi_t\) together with proximal regularization induces hysteresis and smoothing, avoiding churn.

To enhance robustness, the online subproblem retains SOCP relaxations of \emph{chance constraints} using variance bounds \(\sigma^2_{i,\tau}\), \(\varsigma^2_{u,\tau}\):
\begin{align}
\mathrm{Delay}_{i,\tau}(a,r)
+ \vartheta_i\,\sigma_{i,\tau}
&\;\le\; D_i^{\max},
\\
K_{u,\tau}
+\sum_{e\in\mathrm{In}(u)}\hat g_{e,\tau}
-\sum_{i}k_{i,u,\tau}
-\varrho_u\,\varsigma_{u,\tau}
&\;\ge\; 0,
\end{align}
where \(\vartheta_i,\varrho_u\) are set from target confidences to ensure probabilistic feasibility under disturbances. If infeasibility persists, a \emph{feasibility recovery} is triggered by minimizing relaxation magnitudes \(\sum_i \omega_i\zeta_i\) that correspond to reduced reporting, log aggregation, or temporary protection downgrades on low-weight traffic, while preserving hard compliance.

\section{Evaluation Methods}

We evaluate the scheme in a two–timescale simulation: a slow layer for day–ahead/intra–day variability (market rhythms, weather, maintenance) and a fast online layer at minute/second granularity. The platform jointly emulates time–varying QKD key yields, bursty business traffic, and regime switches (normal $\rightarrow$ degraded $\rightarrow$ outage), and reports a unified set of metrics for fair, repeatable comparisons.

\subsubsection{Testbeds and timelines}
We use two representative VPP systems based on the IEEE 33–bus and 123–bus feeders. Each feeder hosts portfolios of PV, wind, batteries, and controllable loads aggregated by a VPP operator. Time is slotted with $\Delta t\in\{1,5\}\,$minutes for the communication/security layer (and sub–second internal queuing if needed); evaluation windows span 1–24 hours to cover diurnal patterns and multiple regime transitions.

\subsubsection{Traffic and message classes}
Five message classes are instantiated to reflect VPP operations (metering, market interaction, dispatch, settlement, audit). Class–specific arrivals follow non–homogeneous Poisson/renewal processes driven by daily load and clearing rhythms, with peak amplifications around market and settlement windows. Payload sizes adhere to industry profiles; class TTLs and importance weights are inherited from the system model (not repeated here).

\subsubsection{QKD overlay and classical backhaul}
We synthesize a metropolitan–scale QKD overlay with 16–24 nodes and 28–40 links over fiber maps; per–link yields vary with weather (QBER/SNR surrogates) and planned outages, creating normal/degraded/outage regimes. Each node maintains a finite TTL key pool with expirations. The classical backhaul is an L3 IP fabric (1–10\,Gbps). We enable three security options (OTP{+}WC, AES{+}WC, AES{+}MAC) with configurable tag lengths and session refresh rates; cross–domain transfer caps and intra–domain quotas enforce administrative boundaries.

\subsubsection{Adversarial and stress scenarios}
To stress robustness without overfitting, we inject “steady–shock–recovery’’ patterns via a hierarchical generator that superposes exogenous triggers (e.g., extreme weather, industry alerts) on a drifting baseline. Attack/query durations are heavy–tailed and synchronized with peak periods; maintenance events create short key–famine windows.

\subsubsection{Comparators and ablations}
We compare against: (i) a static security baseline with fixed strategy maps; (ii) a fixed–priority greedy policy; (iii) a “no–QKD’’ computational–security reference (upper bound on latency when confidentiality is relaxed); and (iv) a clairvoyant oracle (unreachable reference). Ablations remove, one at a time, forecasting, the emergency reserve, degradation (OTP$\!\rightarrow$AES switching), and DRR–style arbitration to quantify marginal contributions.

\subsubsection{Metrics and reporting}
We report (i) latency: per–class P50/P95/P99 and violation frequency vs.\ class deadlines; (ii) reliability: passive timeouts vs.\ active drops; (iii) key/resource efficiency: successful critical messages per key bit, key–pool occupancy/expiry loss, cross–domain key–flow share; and (iv) implementation footprint: per–slot decision latency. Unless stated otherwise, statistics are averaged over 30–100 Monte Carlo runs with fixed seeds; we provide mean and 95\% confidence intervals and release configuration files for reproducibility. Numerical results are presented in the Results section.

\section{Results and Discussions}
\subsection{Overall Performance}
\begin{figure}[t]
  \centering
  \includegraphics[width=\linewidth]{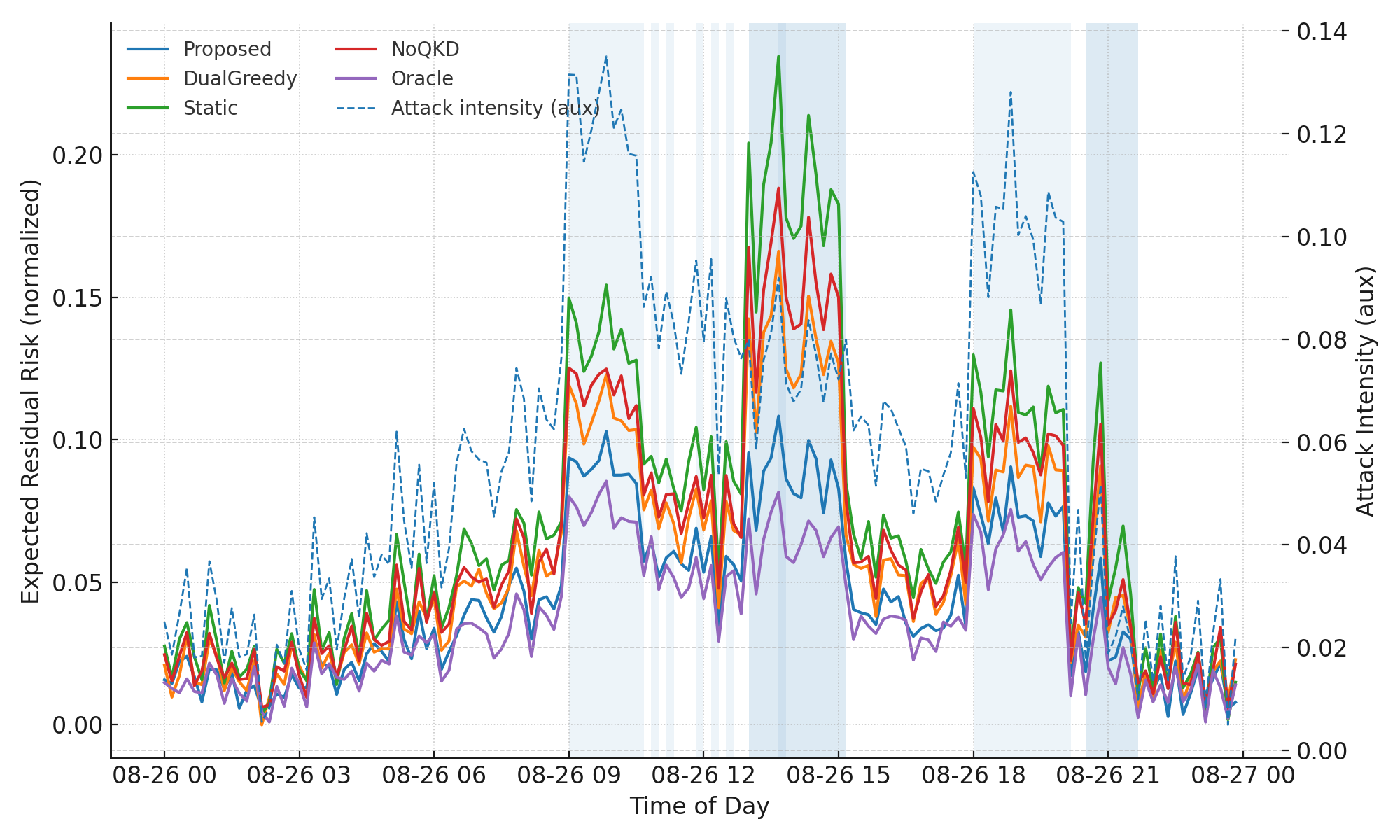}
  \caption{Overall expected residual risk over time for all methods. Shaded bands indicate key-yield shocks and high attack-intensity windows; a twin y-axis overlays the attack intensity series to contextualize spikes.}
  \label{fig:r1-risk-time}
\end{figure}

\begin{figure}[t]
  \centering
  \includegraphics[width=\linewidth]{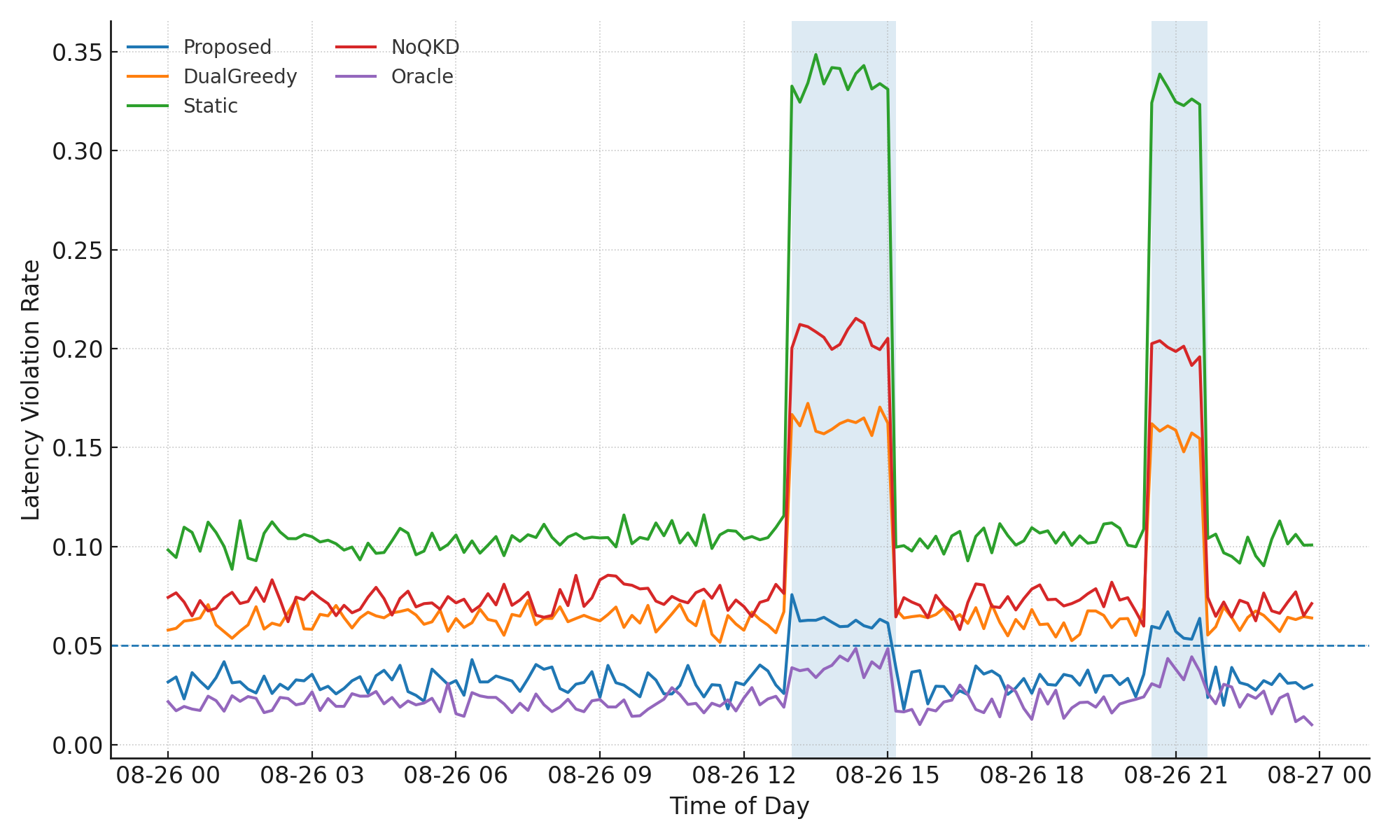}
  \caption{End-to-end latency violation rate over time. Shaded bands denote key-yield shocks; the dashed horizontal line marks an SLA reference (e.g., 5\%).}
  \label{fig:r1-latency-time}
\end{figure}

As shown in Fig.~\ref{fig:r1-risk-time}, the \textit{Proposed} controller tracks the oracle throughout the day while damping spikes in both high-attack and key-yield shock windows (shaded). Relative to dual-greedy and static baselines, it exhibits flatter peaks and faster post-shock decay, consistent with a price--threshold rule that routes scarce keys to high-\(\mathrm{MSV}\) classes exactly when shocks hit. Morning and evening pulses lift risk for all methods, yet the proposed curve stays below no-QKD/static, indicating that hybrid IT/CT with adaptive refresh meaningfully reduces exposure. Latency results in Fig.~\ref{fig:r1-latency-time} mirror this: violations rise system-wide under shocks, but the proposed policy remains near the SLA and re-enters compliance quickly, whereas greedy lingers and static plateaus—evidence that proximal smoothing and incremental updates prevent over-reaction.

\begin{figure}[t]
  \centering
  \includegraphics[width=\linewidth]{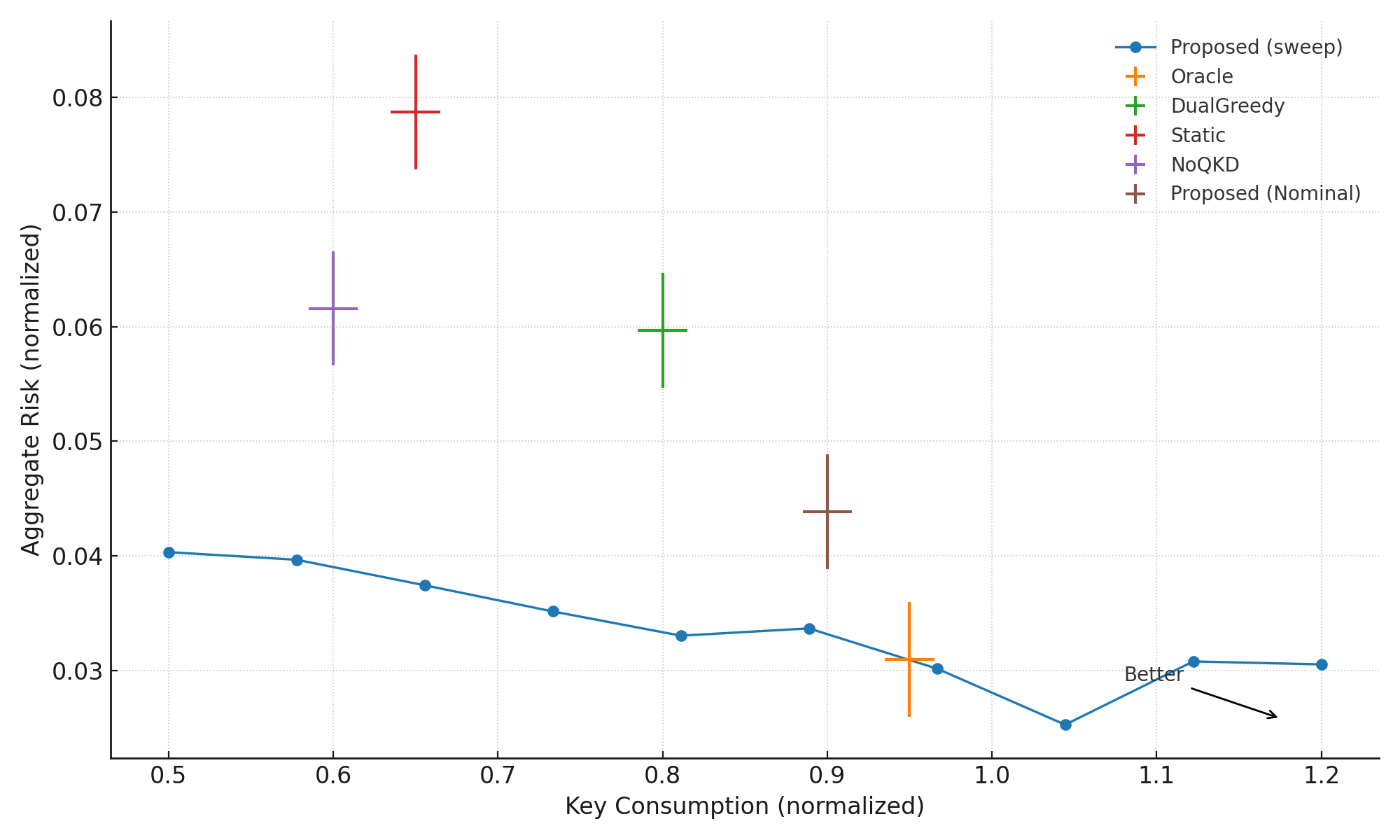}
  \caption{Illustrative risk--key consumption Pareto front. The ``Proposed'' sweep traces a frontier; single points mark comparator policies with small error bars. The arrow indicates the direction of improvement (lower risk with less key use).}
  \label{fig:r1-pareto}
\end{figure}

The risk–key trade-off in Fig.~\ref{fig:r1-pareto} reinforces the advantage: budget sweeps yield an outward-shifted frontier that Pareto-dominates comparators across a broad range, with diminishing returns once the highest-\(\mathrm{MSV}\) traffic is saturated. The no-QKD reference uses fewer quantum keys yet stays off-frontier, underscoring the unique gains from information-theoretic authentication and frequent refresh. Overall, the offline+online design balances residual risk, latency, and key efficiency, remains robust to shocks, and offers interpretable behavior via shadow prices.

\subsection{Resource dynamics and the price--threshold mechanism.}
\begin{figure}[t]
  \centering
  \includegraphics[width=\linewidth]{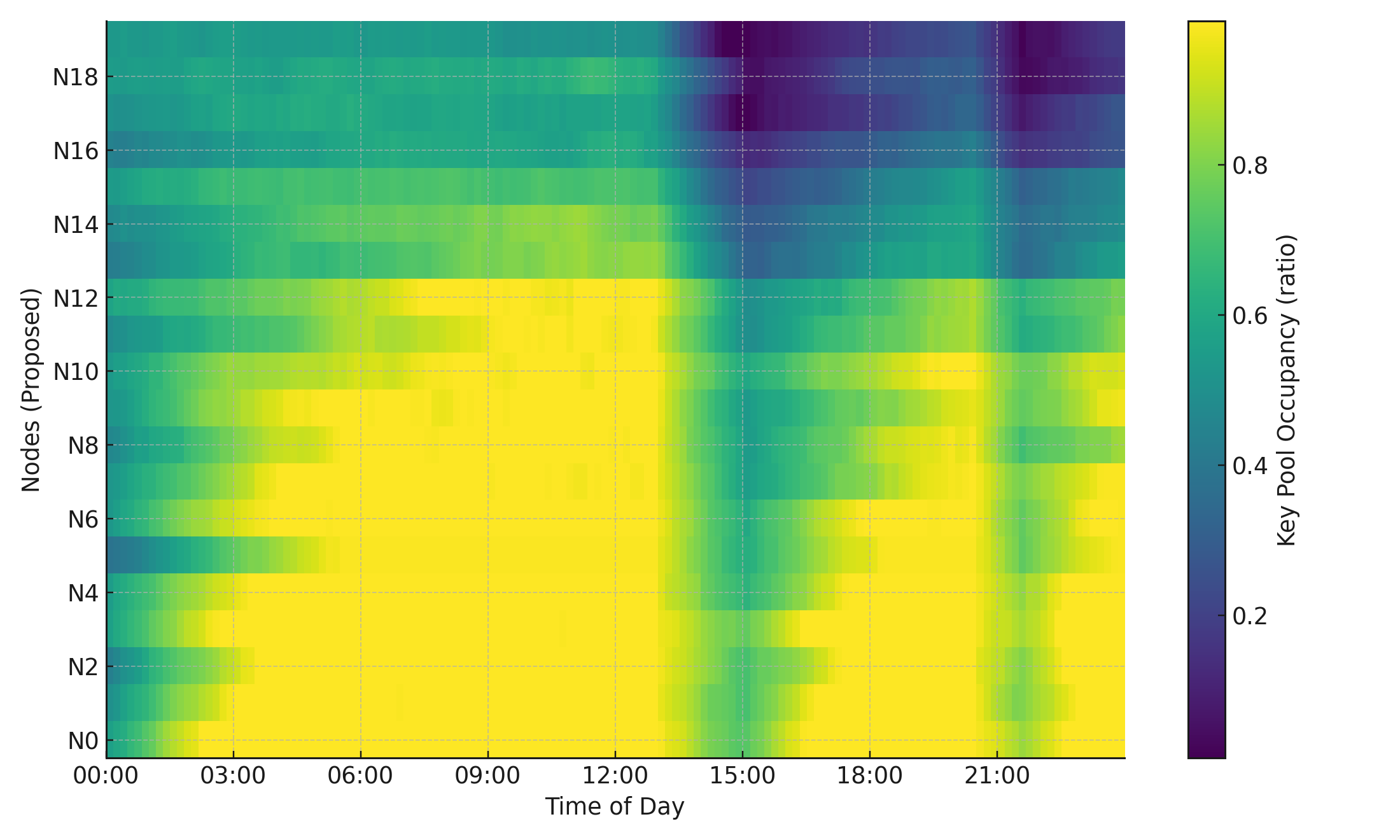}
  \caption{Node-by-time heatmap of key-pool occupancy under the Proposed policy. Darker regions indicate tighter availability during stress, exposing spatial--temporal heterogeneity and bottleneck nodes.}
  \label{fig:r2-heatmap}
\end{figure}

\begin{figure}[t]
  \centering
  \includegraphics[width=\linewidth]{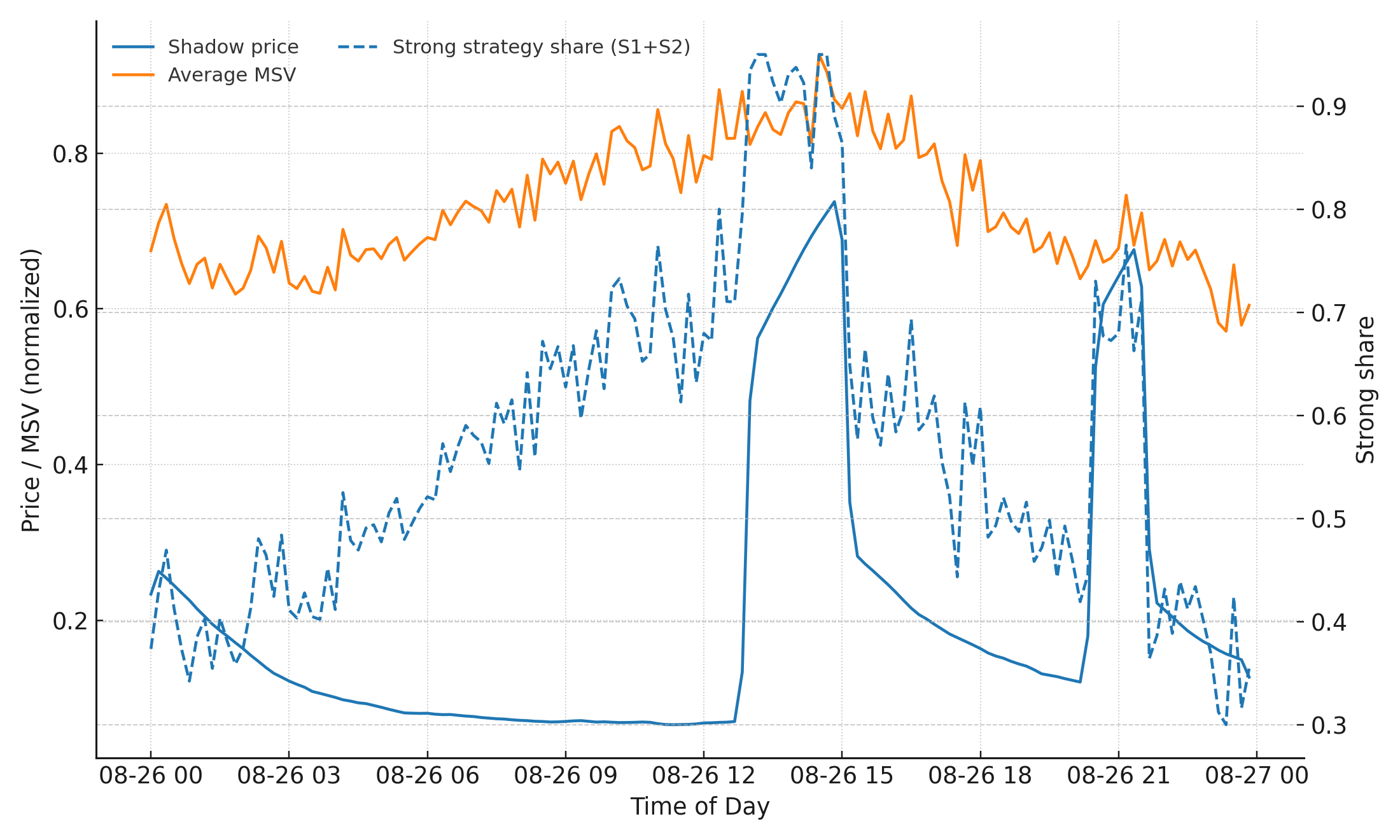}
  \caption{Time series of aggregated shadow price and average marginal security value (left axis), with the share of strong strategies (S1+S2) on the right axis. Peak alignment evidences the price--threshold mechanism and adaptive reallocation.}
  \label{fig:r2-price-msv}
\end{figure}

Figure~\ref{fig:r2-heatmap} shows clear spatio--temporal heterogeneity in key-pool occupancy under the \emph{Proposed} controller: stress windows trigger sharp drawdowns at relay/edge nodes with slow post-shock replenishment (a characteristic ``V''), consistent with short bursts of key spending on high-value traffic. In Figure~\ref{fig:r2-price-msv}, the aggregate shadow price rises in step with the average marginal security value (MSV), while the share of strong strategies (S1+S2) increases precisely during shocks. This co-movement---price, MSV, and strong-share---is the signature of the price--threshold rule: when per-bit security return exceeds the endogenous threshold \(\bar{\pi}\), the controller raises tag length and/or refresh, concentrating scarce keys where risk reduction per bit is largest.

\begin{figure}[t]
  \centering
  \includegraphics[width=\linewidth]{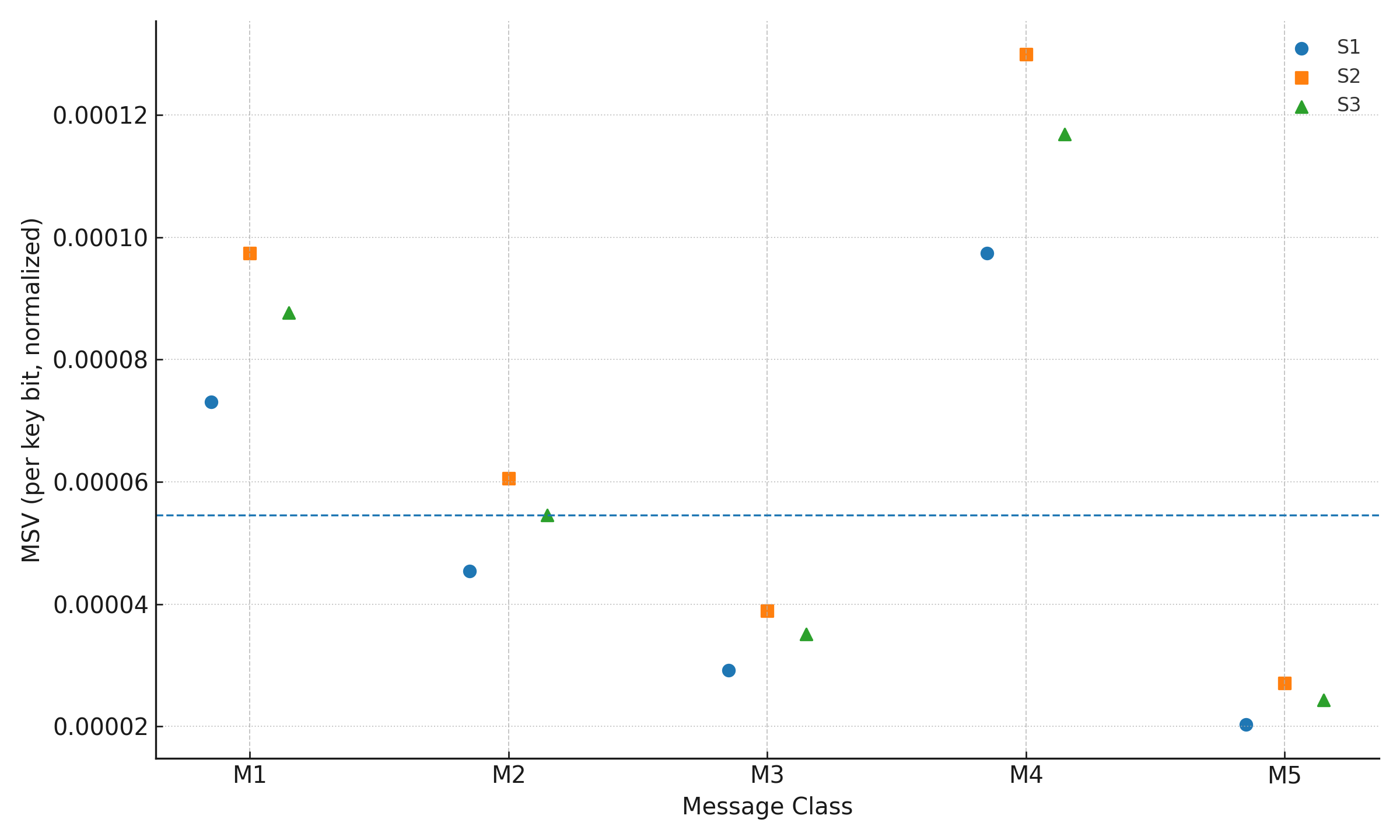}
  \caption{Per-class MSV at a representative high-stress slot with the aggregated shadow-price threshold mapped onto the MSV scale. Points for M1/M4 under S1/S2 predominantly lie above the dashed line, indicating priority hardening, while lower-return configurations remain below.}
  \label{fig:r2-msv-scatter}
\end{figure}

Figure~\ref{fig:r2-msv-scatter} makes the threshold geometry explicit: M1/M4 under S1/S2 sit mostly above the dashed line (priority hardening), whereas many M3/M5 under S3 fall below (lighter protection). After shocks, both occupancy and strong-share revert, showing the policy does not lock into over-protection: as scarcity eases and \(\bar{\pi}\) drops, allocations unload naturally, restoring sustainable key turnover. Overall, the alignment of drawdowns, prices, and strategy shares provides mechanism-level evidence that price--threshold scheduling is interpretable and value-aware, preserving latency while suppressing residual risk under volatile supply and threats.

\subsection{QoSec and latency compliance for key classes (M1 \& M4).}
\begin{figure}[t]
  \centering
  \includegraphics[width=\linewidth]{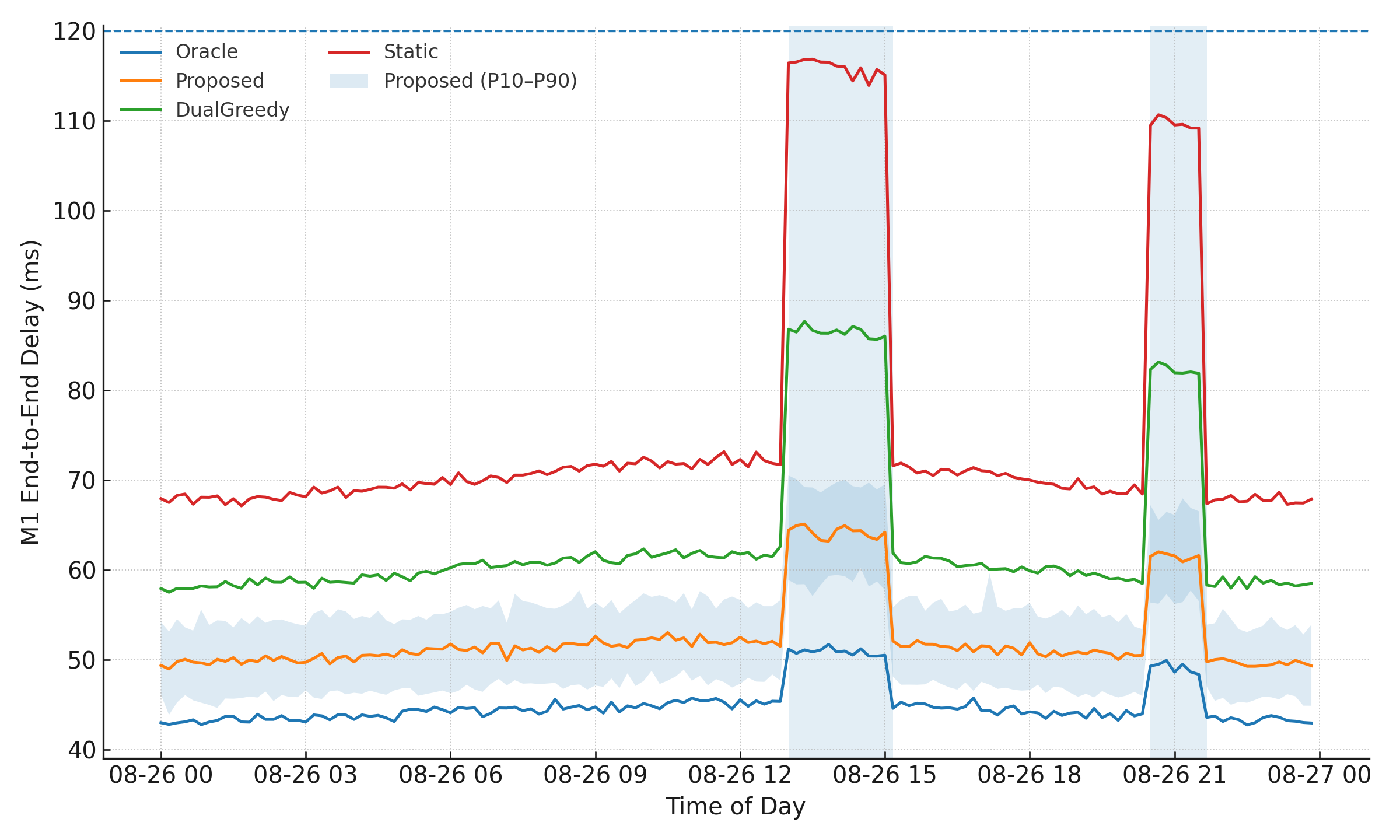}
  \caption{Time-resolved quantiles of M1 end-to-end delay across 80 Monte Carlo replications. Solid lines denote per-method medians; the shaded band shows the P10--P90 envelope for the \textit{Proposed} policy. Dashed vertical line is the SLA (120\,ms); shaded windows indicate stress periods.}
  \label{fig:r3-m1-quantiles}
\end{figure}

\begin{figure}[t]
  \centering
  \includegraphics[width=\linewidth]{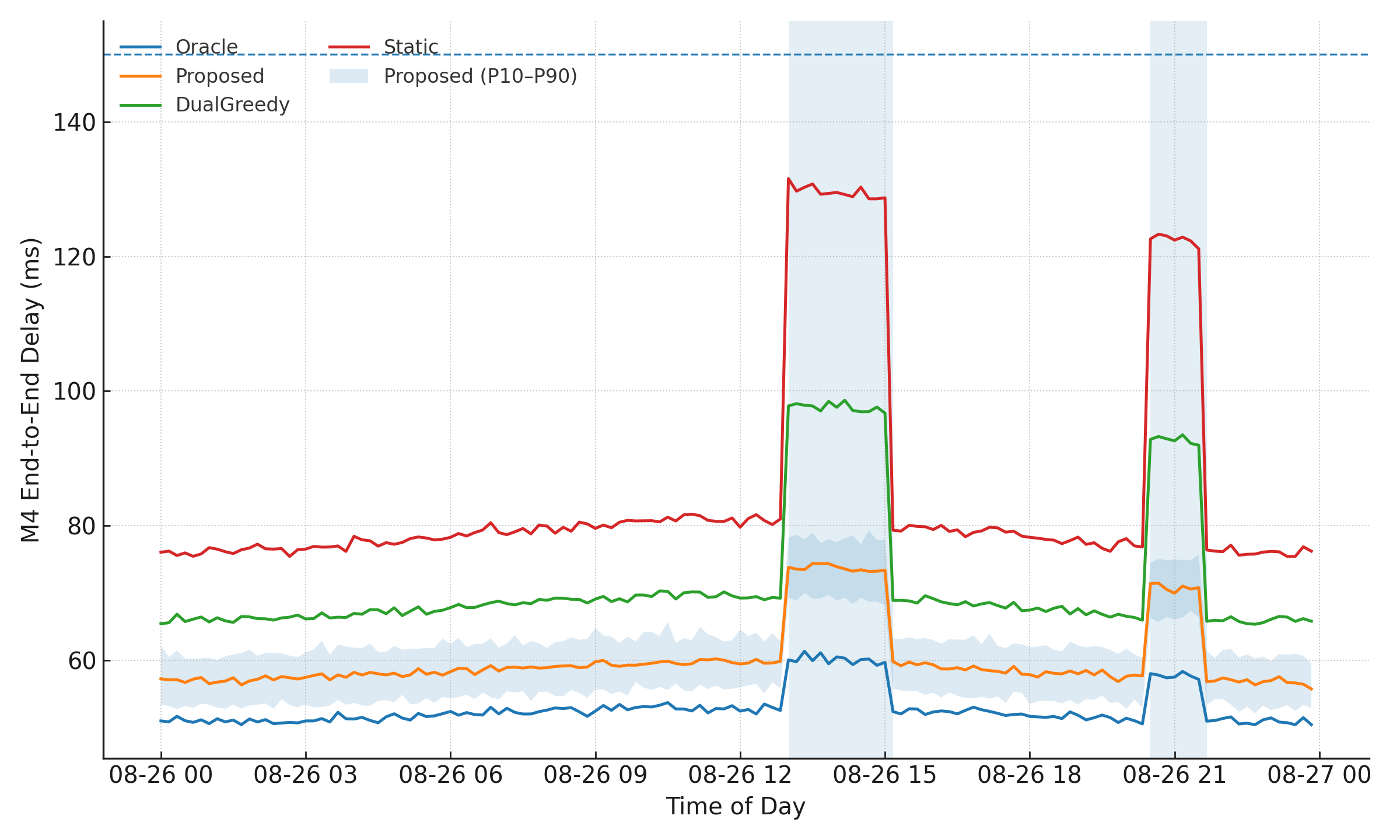}
  \caption{Time-resolved quantiles of M4 end-to-end delay across 80 replications. Solid lines are medians; the shaded band is the P10--P90 envelope for \textit{Proposed}. The SLA is 150\,ms (dashed).}
  \label{fig:r3-m4-quantiles}
\end{figure}

The time-resolved quantiles in Fig.~\ref{fig:r3-m1-quantiles} and Fig.~\ref{fig:r3-m4-quantiles} show that the \textit{Proposed} controller stochastically dominates \textit{DualGreedy} and \textit{Static}: median delays stay below SLA lines and the P10--P90 band remains tight, even in shaded stress windows. Baselines exhibit higher medians and wider spreads during stress, revealing queueing amplification. The \textit{Oracle} curve is leftmost, but the gap to \textit{Proposed} is much smaller than the gap from \textit{Proposed} to the baselines, indicating most deployable gains come from the price--threshold policy.

\section{Conclusion}
This paper presented a quantum-authenticated aggregation and settlement framework for virtual power plants (VPPs), linking QKD key supply and routing with business-layer security strategies through a key-budgeted risk minimization model and hybrid offline--online control.  Experiments on a representative VPP system show that the proposed controller consistently lowers residual risk and SLA violations compared with greedy and static baselines, particularly during attack surges and QKD yield shocks. The price--threshold mechanism was confirmed: shadow prices track marginal security values, and stronger protections (S1/S2) are allocated to critical classes (M1, M4). Delay quantile analysis further indicates stochastic dominance of the proposed method, with QoSec compliance maintained above 99\%.  
Overall, the framework achieves robust reductions in risk and latency violations while improving key efficiency, validating QKD-enabled, risk-aware scheduling as a practical approach for secure VPP operations.

\bibliographystyle{IEEEtran}
\bibliography{refs}


\newpage
\vfill

\end{document}